# Endow a service-oriented architecture by a decisional aspect

**Abstract.** SOA architecture is more and more used in the companies, The importance of the service orientation and its advantages with the information system of the company, confront us to a new challenge. It is primarily to ensure the decisional aspect of the information system of company by adopting Services Orientated Architecture like support architecture. This engineering must ensure on the one hand the flexibility of the information system and on the other hand, avoids the redevelopment of the system by the decisions request. In the actual position, several obstacles force the installation of the SOA within the company. It is basically a question of the lack of method to be implemented to define the services architecture within the information system of the company which takes account the decisional aspect. Moreover, there's no existing works which treat this challenge.

On the basis of these notes, we are interested, in this paper, on the development of a three-dimensional new architecture for the integration of the decisional aspect in SOA architectures, so that they are used perfectly. The proposal takes support, mainly, on the use of a coupling MAS-SOA. To demonstrate the application of our proposition, we use two study cases: project management and Evapo-transpiration.

## 1 Introduction

In a few years, service oriented architecture (SOA) became a major topic for the information systems of company. More than one new technology or method, it is the convergence of several existing approaches, and the emergence of a strong data-processing adhesion of the directions and trade with the same objective. SOA is founded on the construction of reusable and flexible services, neutrals compared to the platform of communication and which correspond to the businesses processes of the company or to the trade dimension.

At present, the Services Orientated Architecture is mainly intended to intervene on the level of: computing system and trade of the companies by presenting their various applications in the form of a set of independent modules able to be made up called services. Nevertheless, in search of agility, the SOA must exceed the technical framework related on data processing and the trade level to touch the decisional level of the company. In this meaning, the true challenge consists in the extending the SOA to the decisional aspect of the company.

Several research tasks were interested on the services identification problem [26] [27] [28] [29] [30]). The majority of this work leads to a pragmatic step whose deci-sion-making aspect is not treated, or doesn't exist. The decisional vision of the company is considered by all the approaches like under process, and it is left as an undeveloped black box.

However, the increasingly extended use of SOA architecture shows that these days, it is not possible any more to continue applying such steps all alone, but always at the end, we must think of the development of decision-making system. The goal of our work, is to open black box of the decisional vision, and to show its components according to an SOA architecture.

The objective of our work is to answer to a finality. To lead to it, we propose to extend The Services Orientated Architecture on the decisional aspect of the company. Thus, the result is a Services Oriented ecosystem which includes services belonging to various levels: the trade level, information and the decisional level. The resulting services guarantee the agility of the company and offers the necessary architecture and infrastructure to adhere to the decision scenarios. We have called this new architecture of company based on the services and integrates the decisional aspect: DECISION MAS-SOA: new architecture for decision based on agents dedicated to Service-Oriented Architectures

The decision aspects are present in many fields and aim to help the decision maker in his task by providing him all the relevant elements for decision making. However, the service oriented architecture adapted in companies do not reflect the reality accurately where various points of view divergent and often conflict must be considered to arrive at a compromise that gave rise to a new dimension: decisional. A major contribution of this paper is the analysis of "how integrate a decision aspect in SOA, in all fields ?"

In this paper, we will present a proposal for a new architecture of decision based on the models. This proposal will be studied on two angles: The first relates to the definition of the agents implied in a decisional system; where we will pass by a state of the art on the decision systems based on service oriented architecture, and then the proposal for a multi-agents system exploiting rules based decisions.

The second angle treats SOA architecture where we will put the point on the various layers which appear at the time of decision making, these can be the technical layers, or the data related on the business processes and the decision-making processes. This architecture will be exploited thereafter in two cases, the first relates to the company it is "the Projects Management", the second is intended for hydraulics it is "the Evapotranspiration". At the end we will make a comparison between our proposal and the work presented in the state of the art.

## 2   Decision aspect and SOA

### 2.1   State of the art

Currently, all the development approaches based on services proposed by research, consider that the adoption of SOA includes only the business and the information system sight of the company. We briefly present these works according to a chronological order.

- Service Oriented Analysis and Design (SOAD) is an approach improved and interdisciplinary of service modeling, suggested by O. Zimmermann (2009) [22], on the basis of existing development processes and notations.

- Service-Oriented Modeling and Architecture (SOMA) illustrates the activities of a modeling method based on services, proposed by Arsanjani (2004) [27]. For the identification and the specification of service, it combines the three analysis approaches, ascending, downward and middle-out.

- Ivanyukovich and Al (2005) [31] propose the adoption of the development process Rational Unified Process (RUP) for the development based on services. In addition of that, they propose also to adapt it so that it can answer certain specific characteristics of this type of environment.

- A conceptual model, presented by Yukyong and Hongran (2006) [32] called M4SOD (Method For Service Oriented Development). The purpose of this model is to formalize the development process SOA. This method put the accent on the phases of identification and realization of the services.

- Rahmani and Al (2006) [33] propose an approach of modeling and designing of systems based on SOA which uses the architecture directed by the models (Model Driven Architecture, MDA).

- Service-Oriented Unified Process (SOUP) [34], a development process intended for the system based on SOA and suggested by Mittal (2006), use the best elements of RUP and XP (Extreme Programming).

- Chaari et al. (2007)[35] proposes the approach the Services Oriented Company which treat the problem of collaboration between companies.

The recent approaches are developed on the base of the SoaML (Service oriented architecture Modeling Language) language. Currently Three equipped methods use this SoaML language for services modeling. The 1st method was proposed by (Amsden, 2010) [26], it was integrated in version 2.9 of the method SOMA (Service-Oriented Modeling and Architecture) proposed by [27] and supported by RSA (Rational Software Architect). The 2nd method is proposed by Casanave [36] and is supported by the tool ModeDriven1. At last, the 3rd method MBDS (Model-Based Development with SoaML) is proposed by Elvesæter [23] and is supported by the modelisation tool Modelio of SOFTEAM1. The authors of these three methods take an active part in the SoaML specification, this results the existence of two alternatives of services modeling in the specification. But there's no work that treat the decision problem.

All the approaches of the services identification presented in the state of the art are limited to the trade levels and information system for the SOA design, there's no approach in the literature that propose a solution for the decisional aspect in this type of architecture.

DSS (Decision Support System) has always been an interesting research topic and researchers have developed models using the new computational tools and techniques. The study of literature reveals that there are no so many works on the decision support systems based on Service Oriented Architecture.

The first work is that proposed by "Xu Liyuan" and "Al" [3], whose title is "a decision support system for a precise irrigation based on the SOA", the help making decision system select the adequate service according to the needs for the users of the different BPMs, and then it set up the model. At the end, it gives the exact instruction to the irrigation. This system is mainly divided into four levels, the presentation layer, the trade layer, the services layer and the data layer.

"Vassilios Vescoukis" [4] proposed an architectural framework for environmental crisis planning and management systems, incorporating data and presentation services, as well as dynamically selected simulation models able to predict future geo space states from real time and static space data.

The proposal made by "Kamran Sartipi" and "al" [17] presented the enhancement of SOA services to incorporate mined-knowledge interoperability as services, along with data interoperability. The application domains include: financial analysis, tourism, insurance, healthcare, and transportation. The application of the provided mined-knowledge at the point of use would boost the accuracy and convenience of decision making by the administrative personnel. At last, the authors proposed a decision support for Electronic Health.

"Villaseñor Herrera" and "al"[18] define their service-oriented control architecture combined with a MAS for flexible control and reconfigurability in factory automation. The elements of the architetucre were briefly reviewed. In particular, the reasons and requirements for integrating a MAS-based DSS into the architecture were analyzed. Three main reasons were identified: service redundancy, service composition, and scheduling; together with three main requirements: semantic time descriptors, control of concurrent service calls, and the modifications to the KB representations. In order to adapt a MAS to WS protocols, it was proposed to create the WS-ACL language. But the author didn't present how the use of a Gateway Agent, which may relieve legacy platforms hosts MAS from implementing WS protocols.

"Christoph Becker" and "al" [19] described the basic architecture and features of a decision support system for preservation planning based on a service oriented approach for distributed preservation solutions.

"Michael Gebhart" and "al" [21] introduced an approach to determine the impact of design decisions on the characteristics of services in order to support making them. To illustrate their approach, services of a service-oriented surveillance system are designed bearing the surveillance system N.E.S.T. of the Fraunhofer Institute of Optronics, System Technologies and Image Exploitation.

"Olaf Zimmermann" [22] proposed an Architectural Decision Modeling Framework for Service-Oriented Architecture Design, they called it SOA Decision Modeling (SOAD) framework.

Last work is that of "Neil Wheeler" [02], and which consists in proposing a Service Oriented Architecture for an assistance system with the decision of treatment of fuels inter-agencies (Interagency Fuels Treatment Decision Support System (IFT-

DSS)). This last system was proposed by "Funk" and "al". [15], and "Neil Wheeler" proposes an improved architecture of the system of decision containing SOA.

### 2.2 Background and Motivation

All the work presented in the state of the art, relates the application of the Decision support systems based on SOA architecture to a well defined field (irrigation, Health, treatment of fuels inter-agencies, preservation planning, factory automation, software architecture …etc. ). All of them only cover a subset of the method decision, i.e., they focus on the decision problem and put less emphasis on intelligence, design, and choice.

SOA defines a set of concepts organized according to two points of view [23]: business, focusing on the features and the requirements of the business in which the IS will be built, and the information system view, concentrating on the functionalities and processes that need to be implemented in the IS in order to satisfy the business requirements. The importance of our work is to give a third view for the SOA architectures, which is the decision view that can be used in any field according to IDC SIMON [20] decision model.

Multi Agent Systems (MAS), a term used to describe the incorporation of multiple types of agents into various systems, is a way of designing and implementing a system with the advantages of agent entities. We choose to use agents as a decision support tool for use in a Retail DSS. Since the DSS is crucial to the success of most companies, and since we see a potential major role for agents in the business process management MAS seems a likely choice for decision support architecture, then implementing the system for SOA use.

In this article we propose an architecture supported by the MAS, dedicated to the decisions based on the models, in any field (undertaken, chemistry, physics, hydraulics… etc.).

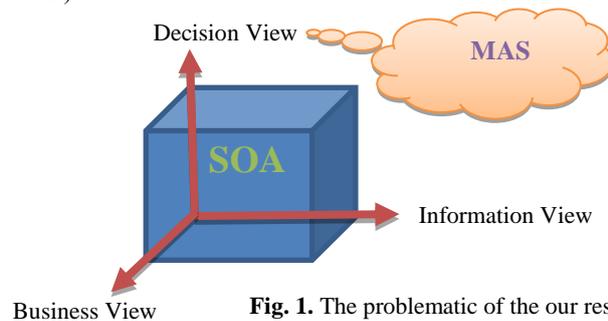

**Fig. 1.** The problematic of the our research

## 3 SUGGESTED ARCHITECTURE

The originality of our architecture is due to the simultaneous use of a multi agents system (MAS) and service orientated architecture (SOA) to endow this last by a decisionnal aspect. The literature offers few examples of coupling these two types of rep-resentations of reality [18]. Our architecture is based on the MAS to propose an innovating system which deals with the decision-making process. The use of MAS is justified by its autonomy, the capacity of negotiation and self-organization for decision making. Moreover emergence of the intelligence results from the interactions between the agents to define their own problem resolution scenario.

### 3.1 Suggested MAS architecture

Multi Agents Technology already proved reliability in many fields by their capacity of modeling, they make it possible to represent the interactions between various entities being able to cooperate, negotiate and communicate. The speakers of the system, whom we study, are the various decision makers or experts who have their own objectives. That implies that the decision-making process is distributed between the various entities implied by this decision of group. MAS module will have the role of representing the various actors who have their own objectives and preferences. In order to face this decision of group where various points of view must be taken into account, it is essential to pass by a phase of communication to arrive at a beneficial consensus at the groups. For this purpose, we equip MAS module with a communication protocol, putting in scene a supervisory agent, an Editor agent and Arguer agent.

**The Agents Modeling.** Our modeling agent is based on the methodology Aalaadin [4], which is based on the agent concepts, groups and role to define a real organization.

- An agent is defined as being an autonomous and communicating entity playing of the roles within various groups;
- A group is composed of various agents;
- A role represents a function, a service or an identification of an agent pertaining to a particular group. In our work, there are three types of agent roles: the Supervisor, the Editor and the Arguer;

  - *The Supervisory agent*: is responsible for the good progress of Process trade, thus the Decision-making process. This agent indicates all the anomalies in the course of the two processes.
  - *The Editor agent*: makes it possible to edit the decisional models, to check the validity of the latter, and to record the models, the indicators and the indices.
  - *The Arguer agent*: is the agent concerned by the decision, the role of this agent is to seek the adequate services to find the indicators, moreover, to present the decision indicators according to a mode of visualization (Gauge, Text, Histograms….etc.) in the convenient moment.

*Functioning*. The decision maker introduces the indices and the decision models which must be defined and checked by the Editor agent (Fig.2). In the case of error, this last sends a message to the Supervisory agent. The indices will be recorded in the base of indices like services of low granularity (like trade services), in the same way, the models will be validated by the Editor agent, then recorded in the base of Models like services of average granularity (engineering service (Functional)), they are the high level services that mask the indices service to the composite applications. In the case of error the agent sends a message to the Supervisory agent.

At the level of the models introduction, the Editor agent identifies the indicators of the system, and it records in the base of the indicators like services of large granularity (Applicative service).

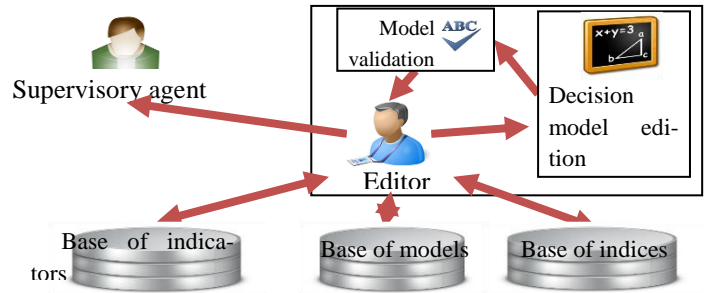

**Fig. 2.** The role of the Editor agent

The purpose of the arguer agent is to calculate the indicators requested from the decision, and to post them according to a mode of visualization (gauge, text, histogram… etc) to help the decision maker by a better view of the value and the importance of these indicators.

When the user selects the indicators that he needs, and the mode of visualization which is appropriate to him, this operation allows the arguer to launch out, this last calls the trades services, indices services and model services to have the indicators requested by the decision maker (fig.3).

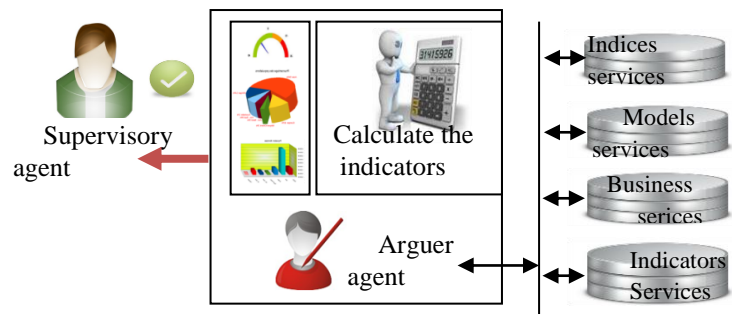

**Fig. 3.** The role of the arguer agent

### 3.2 SOA Architecture.

Our Decision-MAS-SOA architecture illustrated by (fig.4) is made up of four layers in accordance with the IBM model [6] and the architecture of [7], [8] and [9]. The SOA-Decision-MAS must make it possible to define specific indices to each company and to build indicators adapted to the needs for the company. An abstract view of our contribution is presented in (fig.4). This figure shows a Service Oriented Architecture based on layers with different provision levels. Each layer has a specific functionality which is described as follow:

- The Layer 1 "*Data*": it contains two under layers; the first is "Trade" which includes the services trade (CRUD) trade process carried out in the company. The second under layer is the "Indices", its goal is to safeguard the indices services. The Editor agent intervenes on the level of this under layer, to guarantee the recording of these services.

- The Layer 2 "*Technique*": it contains two under layers: under layer "Function" which represents the function services of the process trade. Under layer "Models", which included model services recorded by the Editor agent.

- The Layer 3 "*Action*": it contains two under layers: Applicative and Indicators. The first gathers the Applicative services of the process trade, and the second ensures the appreciation of the indicators, using the arguer agent, and a mode of visualization.

- The Layer 4 "*Presentation*": it contains interfaces, and ensures the communication between the user and the system, that is to say to carry out the process trade, or to make a decision. The Supervisory agent intervenes in this layer.

Moreover, we conceived our approach so that it respects IDC model of SIMON [20]. We have followed what is perhaps the most widely accepted categorization of the decision-making process first introduced by Herbert Simon. Simon's categorization of the decision-making process consists of three phases (IDC): Intelligence, Design, and Choice.

On the level of the presentation layer, the decision maker can make a decision according to the indicators, in more the mode of visualization helps the decision maker in his choice. In the model layer, the platform makes it possible to conceive the solution (indicators). The purpose of the last layer "indices" is to identify and formulate the problem.

The goal of separation between under layers is to apply the solution in fields of decision out the company (Chemistry, Physique, Medicine… etc), i.e., in cases where we do not have a process trade to carry out, we have only one decision-making process based on rules.

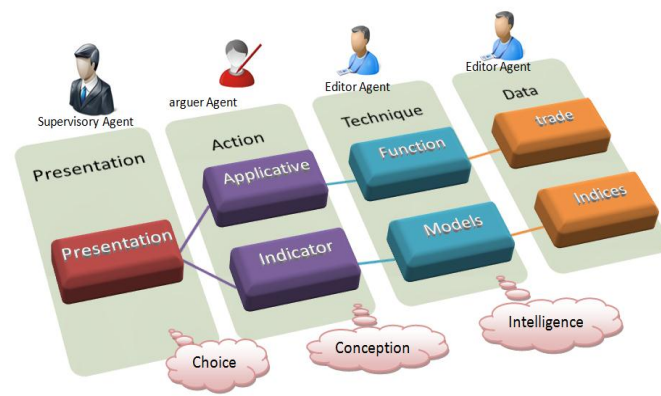

**Fig. 4.** SOA-Decision-MAS architecture

### 3.3 Contribution

The Decision-MAS-SOA architecture assists the decision maker, without replacing him, during the decision-making process. It makes it possible to the decision maker to have access to many knowledge, to synthesize them and test various possible choices.

The architecture below shows also the flexibility of this architecture, because on the one hand one can make decisions on the level of the company, for this reason we kept the layer "trade, Function, Process" defined in SOA architectures of a company, and on the other hand, one can make decision out the company, it is in the fields of: Chemistry, physics, Mathematics, Medicine… etc to get this goal, we separated the layers which define the company (Trade, Organization and Process)from the layers which show the decision (Indicator, Model, index, Mode of visualization).

## 4 Validation of the suggested architecture

This part has as an ambition to validate the conceived architecture. We will apply the present architecture to two cases, the first contains a business process i.e. in the company, it is the project management. The second, it is an application out the company, it is in the hydraulic field, this case does not contain a process trade, but it contains only decision models.

### 4.1 The architecture applied in the Project management

The project management or project control is a step aiming to structure, ensure and optimize the good progress of a sufficiently complex project in order to[25]:
- To be planned in time: it is the object of planning,
- To be budgeted,
- To control the risks,
- To reach the level of quality wished.

The principal role of Decision aspect is to control a project effectively, to reach this objective, we use earned value analysis method to design our DSS.

Earned value analysis (EVA) [24] is a method of performance measurement. Many project managers manage their project performance by comparing planned to actual results. With this method, one could easily be on time but overspend according to the plan. A better method is earned value because it integrates cost, schedule and scope and can be used to forecast future performance and project completion dates. It is an "early warning" program/project management tool that enables managers to identify and control problems before they become insurmountable. It allows projects to be managed better on time, on budget. Following is the summary of important Earned Value terms and formula.

The application of architecture implies the determination of indicators of piloting of the project which are tools of navigation and decision. They make it possible to measure a situation or a risk, to give an alarm or contrary to meaning the correct advance of the project. An example on the indices, models and indicators of system is represented by the following table:

**Table 1.** Earned Value Management Terms

| Term | Description | Interpretation |
|---|---|---|
| PV (BCWS) | Planned Value | What is the estimated value of the work planned to be done? |
| EV (BCWP) | Earned Value | What is the estimated value of the work actually accomplished? |
| AC (ACWP) | Actual Cost | What is the actual cost incurred? |
| BAC | Budget at Completion | How much did you BUDGET for the TOTAL JOB? |
| EAC | Estimate at Completion | What do we currently expect the TOTAL project cost? |
| ETC | Estimate to Complete | From this point on, how much MORE do we expect it to cost to finish the job? |
| VAC | Variance at Completion | How much over or under budget do we expect to be? |

**Table 2.** Earned Value Management Formula and Interpretation

| Name | Formula | Interpretation |
|---|---|---|
| Cost Variance (CV) | EV – AC | NEGATIVE is over budget, POSITIVE is under budget |
| Schedule Variance (SV) | EV – PV | NEGATIVE is behind schedule, POSITIVE is ahead of schedule |
| Cost Performance Index (CPI) | EV / AC | I am [only] getting ______ vents out of every $1. |
| Schedule Performance Index (SPI) | EV / PV | I am [only] processing at ______ % of the rate originally planned. |
| Estimate At Completion (EAC)  **Note** : There are many ways to calculate EAC. | BAC / CPI  AC + ETC  AC + BAC – EV  AC + (BAC – EV) / CPI | As of now how much do we expect the total project to cost $ _____.  • Used if no variances from the BAC have occurred  • Actual plus a new estimate for remaining work. Used when original estimate was fundamentally flawed.  • Actual to date plus remaining budget. Used when current variances are atypical.  • Actual to date plus remaining budget modified by performance. When current variances are typical. |
| Estimate To Complete (ETC) | EAC – AC | How much more will the project cost? |
| Variance At Completion (VAC) | BAC – EAC | How much over budget will we be at the end of the project? |

The next figure shows the modeling of Project management business process with BPMN (Business Process Modeling Notation).

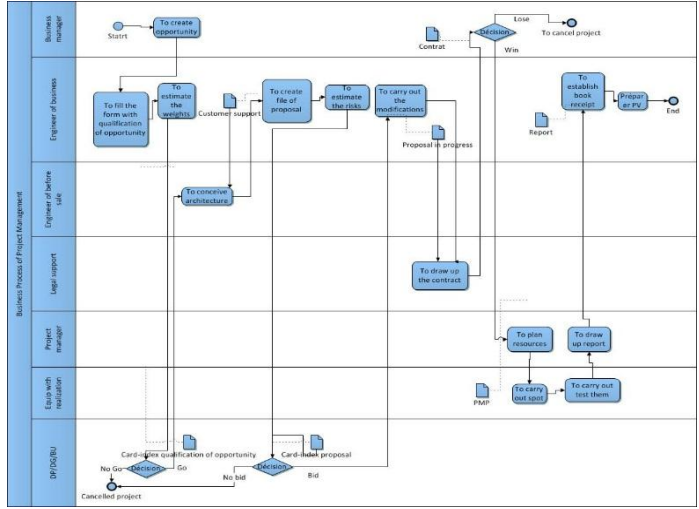

Fig. 5. BPMN of Project Management

The following figure shows the Service Architecture diagram of SoaML.

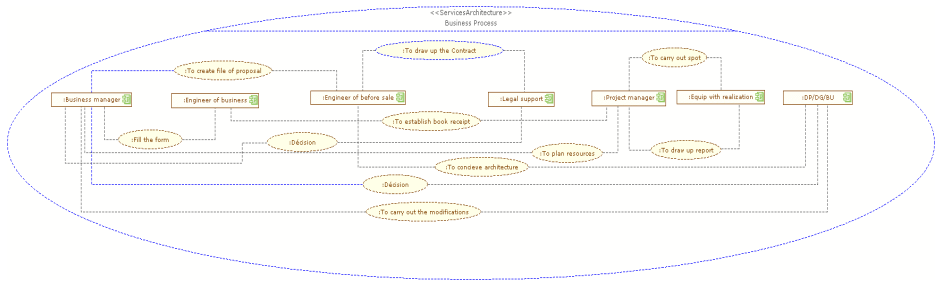

Fig. 6. Service Architecture diagram of Project Management.

The next figure explains how to apply architecture to the project management.

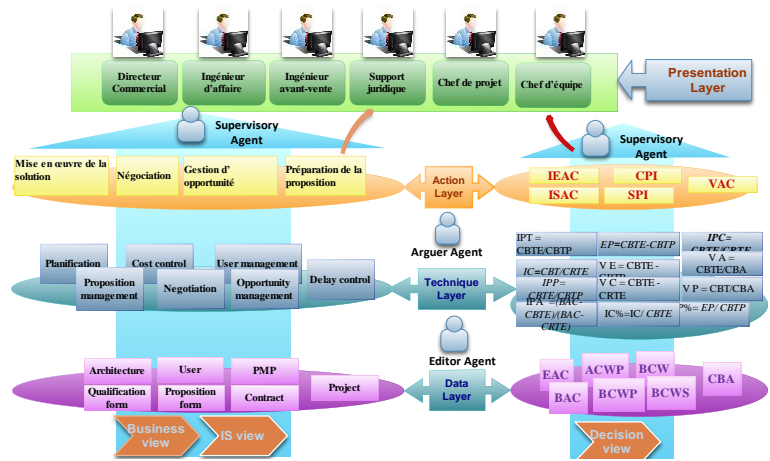

**Fig. 7.** Application of the *DECISION- MAS-SOA* to Project Management

**4.2 The architecture applied in the "hydraulic" field**

We applied architecture to calculate the evapotranspiration. The latter is an essential element to an adequate management of water on a catchment area. The variation of the evapotranspiration can, in a significant way to influence the climate, by varying flows at the border ground-seedling-atmosphere. For example, evaporation can be limited by the quantity of water contained by the ground, i.e. its moisture. By taking account of the importance of this variable in calculating the water assessment.

We apply architecture to have the evapotranspiration of reference by various mathematical models, according to types of data which the user can have at his disposal.

According to the Turkish model which is presented in the (fig.8).

*The indices* are : decadal total solar radiation, T: average temperature of the period considered, Ra: extraterrestrial radiation (calcm-2 J), NR: possible astronomical duration of insolation (hour/month or decade), N: duration of effective insolation (hour/month or decade). Calculates is done for a moderated climate.

*The indicators* are : Rs (solar radiation) Day laborer, Rs (solar radiation) decadal, Decadal temperature, ET (potential evapotranspiration) decadal, ET (potential evapotranspiration) monthly.

*Models examples:*

$$[ET]\_mensuel = 0{,}4.(R\_s + 50).t/(t+15)$$
$$[ET]\_dic = 0{,}13.(R\_s + 50).t/(t+15)$$
$$R\_s = R\_a.(a + b.n/N)\ ;\ d\_r = 1 + 0{,}033.\cos((2.\pi)/365.J)$$
$$\delta = 0{,}409.\sin((2.\pi)/365.J - 1{,}39)\ ;$$
$$\omega\_s = arccos.[-\tan(\emptyset).\tan(\delta)]$$
$$\emptyset(rad) = \emptyset(lat).\pi/180$$

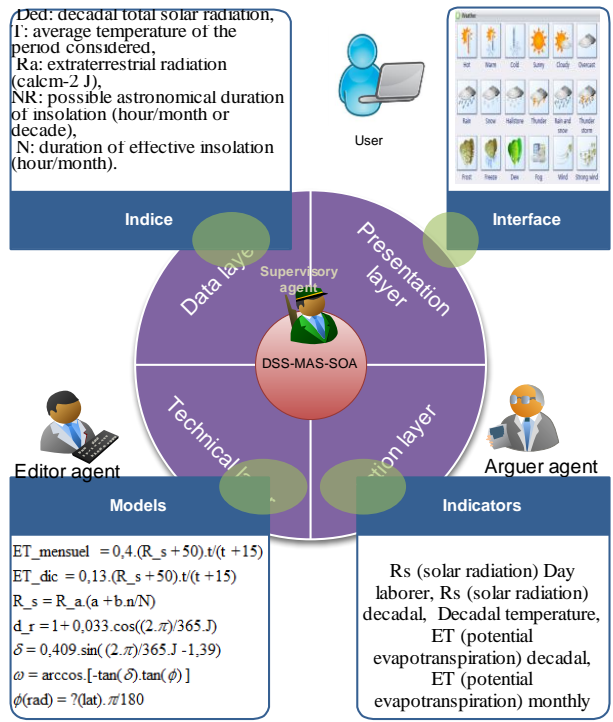

**Fig. 8.** Application of the *DECISION- MAS-SOA* to the hydraulic field

### 4.3 Comparison between SOA architectures intended to make decisions

The following table shows a comparison between our suggested architecture and the work presented in the state of the art.

**Table 3.** comparison between our suggested architecture and the work presented in the state of the art

| | Field | Technique of decision | Solution SOA | Data used | The process of decision | Benefit |
|---|---|---|---|---|---|---|
| **Neil [02]** | treatment of fuels inter-agencies | Not | Divided into four levels defined by the author | Geo-space Real | Is not defined | Decisional dimension |
| **Xu Liyuan [3]** | Irrigation | Not | Divided into four levels defined by the author | Real | Is not defined | -Decisional dimension -Business dimension |
| **Kamran Sartipi and al [17]** | Electronic Health | Not | Divided into four levels defined by the author | Real | Is not defined | -Decisional dimension -Business dimension |
| **V. Villaseñor Herrera and al [18]** | Industrial Automation | MAS | Divided into four levels defined by the author | Real | Is not defined | Decisional dimension |
| **Vassilios [4]** | environment of information management | Not | Divided into four levels defined by the author | Geo-space Real | Is not defined | Decisional dimension |
| **Christoph Becker and al [19]** | preservation planning | Not | Is not defined | Is not defined | Is not defined | Decisional dimension |
| **Michael Gebhart and al [21]** | characteristics of services | Not | Is not defined | Is not defined | Is not defined | Decisional dimension |
| **Olaf Zimmermann [22]** | software architecture | Not | IBM model | Is not defined | Is not defined | Decisional dimension |
| **decision- MAS - SOA** | **Company (Management project, stock… etc), Hydraulics (the evapotranspiration), Physics, Chemistry… etc** | **MAS** | **Divided into four levels conform to other SOA standardized architectures** | **Real** | **Architecture follows model IDC of SIMON** | **Decisional dimension Business Dimension IS dimension** |

## 5   Conclusions and Outlines

At the end of this article, we proposed an SOA decisional architecture based on one coupling MAS- SOA likely to bring an effective help to the decision makers.

In this article, we proposed the interest of the decision-making aid like a new way of design of applications of the oriented service approach, while focusing us on models.

In this work, a service oriented architecture and multi-agents allowing the resolution of problem of decision-making aid were presented. It is made up of a user interface, Action layers, Technique and of a Data layer, each layer is made up of two under layers, to represent the process trade, other intended for the decision-making process. In each level we defined an agent responsible for a whole of the tasks, in order to ensure the course of decision-making process. These various layers allow to the user to carry out his process trade, and solve his problem according to his needs. The suggested architecture assists the user throughout his decision-making process without substituting him.

Thus, we proposed the definition of new service types intended to carry out the decision-making aid. Three service types were proposed:

- Services of indices, intended to represent the general framework, the context in which the decision is carried out,
- Services of model, allowing to identify the rules of decision,
- Services of indicator, intended to indicate the waiting of the decision maker.

Thus, we showed as illustration, two applications of the proposal to illustrate the variety of use of our architecture.

In our future work, we envisage the enrichment of our architecture (MAS-SOA) to which we will add new modules and new classes which will allow to model the real systems more easily and to develop other computing systems of decision-making aid.

Like prospect, there remains to us the implementation of a framework suggested architecture, and to test it on different fields. We also think to use another methods and tools of decision-making aid like the multi criterion methods.